# QUANTUM MECHANICS WITH CHAOS:
## Correspondence Principle, Measurement and Complexity

**Andrei P. Kirilyuk**[*]

Institute of Metal Physics, Kiev, Ukraine 252142

chao-dyn/9510013 v3   25 Mar 1999

**Abstract.** The true dynamical randomness is obtained as a natural fundamental property of deterministic quantum systems. It provides quantum chaos passing to the classical dynamical chaos under the ordinary semiclassical transition, which extends the correspondence principle to chaotic systems. In return one should accept the modified form of quantum formalism (presented by the Schrödinger equation) which, however, does not contradict the ordinary form and the main postulates of quantum mechanics. It introduces the principle of the fundamental dynamic multivaluedness (redundance) extending the quantum paradigm to complex dynamical behaviour. Moreover, a causal solution to the well-known problems of the foundation of quantum mechanics, those of quantum indeterminacy and wave reduction, is also found using the same method. The concept of the fundamental dynamic uncertainty thus established is universal in character and provides a unified scheme of the complete description of arbitrary complex system of any origin (physics/9806002). This scheme incorporates, in particular, universal definitions of complexity, (non)integrability, and general solution, as well as the physically complete notion of probability. One obtains thus a self-consistent hierarchic picture of the world characterised by a (high) non-zero complexity and containing the intrinsic causal randomness, where the causally extended quantum mechanics can be consistently interpreted as several lowest levels of complexity (quant-ph/990215,16).

This is a descriptive representation of a large work issued in the form of preprint (IMP 95-1) and now also presented as a series of e-prints (quant-ph/9511034-38) and included into a book by the same author (see the Comments field of this e-print). This presentation contains the table of contents of the full paper (the preprint), its extended abstract, and some sample parts slightly adjusted for convenience.

---

[*] Address for correspondence: Post Box 115, Kiev - 30, Ukraine 252030.
E-mail address: kiril@metfiz.freenet.kiev.ua

# TABLE OF CONTENTS (of the full paper)

# BRIEF CONTENTS (of the full paper)

**Part I.** The effective (optical) potential method is generalised and adapted to obtain a non-perturbative self-consistent description of quantum chaos in Hamiltonian systems. The method provides a reformulation of the Schrödinger equation revealing the multivaluedness of the effective Hamiltonian, i. e. its splitting into many branches, called 'realisations'. Each realisation incorporates the normal complete set of eigenfunctions and eigenvalues for the entire problem. It means that one can never observe more than one realisation at a time. Then chaos appears naturally as noise-assisted 'spontaneous transitions' between the different realisations.

In the low-noise limit, a chaotic system can stay moving in one of the realisations, chosen at random, for a very long time, and this is a manifestation of the "quantum suppression of chaos". However, for an ensemble of systems, or experimental runs for the same system, one will have, even in this limit, many different realisations "populated" with a priori equal probabilities.

Higher noise levels provoke faster change of realisations, the latter preserving their purely dynamical origin. This case corresponds to the intuitively more obvious picture of chaos. In particular, it occurs in the quasiclassical limit for a quantum chaotic system, when the separation between realisations becomes vanishingly small, increasing the relative magnitude of noise, and quantum chaos passes naturally to classical chaos, in full agreement with the conventional correspondence principle. Indeed, by performing the usual quasiclassical transition in our quantum-mechanical results for the periodically perturbed Hamiltonian system, we reproduce such important and well-known features of classical chaos for that system as the transition global chaos - global regularity, the point of this transition, and the asymptotically weak chaos within global regularity.

This introduces the postulate of the fundamental dynamic uncertainty independent of the quantum-mechanical indeterminacy and involving the new version of the Schrödinger equation. The modified formulation of quantum mechanics does not contradict the ordinary one but rather extends it to the case of chaotic dynamics. It provides the 'true' chaos with the intrinsic randomness appearing as a fundamental property of deterministic dynamical systems not reduced to some kind of sophisticated but predictable behaviour. Correspondingly, any reasonably defined complexity, including algorithmic complexity, is greater than zero for quantum chaotic systems, within this scheme.

The fractal character of quantum chaos is naturally deduced within the same method from the modified Schrödinger equation. It is reduced to the fractal properties of the effective dynamical function in question (effective potential) called the fundamental dynamical fractal. The multitude of branches of this object represents the dynamic complexity of the system, while the involved structure of each branch corresponds to fractal properties of the observables. In this way, the full geometrisation of complex dynamics is attained: the system behaviour, whatever complex, can be expressed, in principle by analytical means, through the geometrical characteristics of the fundamental dynamical fractal.

The method of analysis and the mechanism of chaos proposed seem to have very general character, and are not limited either to quantum, or to Hamiltonian systems. Within the concept of the fundamental dynamic uncertainty, any kind of chaos can be expressed by the multivaluedness of the effective dynamical functions leading to instability because of the redundant choice of solutions. Quantum chaotic systems represent a particular case of this intrinsic randomness, which occurs even when they are far from the semiclassical limit, even though partial "quantum suppression of chaos" does exist and is specified in our description, as well as other particular chaotic regimes and features.

The method is presented in detail for the Hamiltonian system with periodic (not small) perturbation, both in its time-independent and time-dependent versions. Illustrations are given for the possibility of its application to real physical systems. The generalisation to other cases of complex behaviour is outlined as well as the issuing universal and unambiguous definitions of the fundamental dynamic uncertainty, randomness, probability, complexity, (non)integrability, and general solution, practically applicable to real dynamical systems of any origin.

**Part II.** The concept of the fundamental dynamic uncertainty, based on the method of the effective dynamical functions and used to re-establish the correspondence principle for the chaotic Hamiltonian systems (Part I), provides also a causal description for the basic properties of quantum measurement, — quantum indeterminacy and wave reduction.

The modified Schrödinger formalism involving multivalued effective dynamical functions reveals the dynamical origin of quantum indeterminacy as the intrinsic nonlinear instability in the combined quantum system of the measured object interacting with the instrument. As a result of this instability, the originally wide measured wave dynamically "shrinks" around a random accessible point of the combined configurational space losing its coherence with respect to other possibilities. We do not use any assumptions on particular "classical", "macroscopic", "stochastic", etc. nature of the instrument or environment: full quantum indeterminacy dynamically appears already in interaction between two microscopic quantum deterministic systems, the object and the instrument, possessing just a few degrees of freedom a part of which, belonging to the instrument, should correspond to local excitations into an open configurational manifold. This dynamically indeterminate wave reduction occurs in agreement with the postulates of the conventional quantum mechanics, including the rule of probabilities, thus effectively transforming them into consequences of the fundamental dynamic uncertainty.

The obtained general results on the dynamically chaotic nature of quantum indeterminacy and wave reduction in the abstract measurement process are illustrated by an example of particular quantum measurement, the canonical projectile position measurement. Using a simple gedanken scheme of the two-slit type we obtain explicitly the dynamically random configurational shrinking of the measured wave function as a manifestation of its effectively nonlinear behaviour during interaction with the instrument. The particular model considered contains just a few non-stochastic degrees of freedom.

The physical consequences of the analysis performed are summarised within a tentative scheme of the complete quantum (wave) mechanics called quantum field mechanics and complementing the original de Broglie ideas by the dynamic complexity concept. It includes the obtained formally complete description at the level of the average wave function of Schrödinger type showing dynamically chaotic behaviour, and in particular, causal quantum indeterminacy and wave reduction. This level is only an approximation, though rather perfect and usually acceptable, to a lower, and the last, level containing the unreduced, essentially nonlinear de Broglie double solution. The latter describes the state of a nonlinear material field and includes the unstable soliton-like high-intensity "hump" moving chaotically within the embedding smooth wave (see also e-prints quant-ph/990215,16). The involvement of chaos, understood within the same concept of the fundamental dynamic uncertainty, provides, at this lower level, "hidden thermodynamics" of de Broglie (for an isolated particle) without, however, the necessity for any real "hidden thermostat" with the corresponding "subquantum level" of reality. The indeterminate reduction of the "piloting" Schrödinger wave during measurement, at the higher level of description, conforms with the chaotic motion of the virtual soliton.

# 1. Introduction (≈ section 1 of the full paper)

The recent emergence and development of the dynamical chaos concept have engendered profound changes in our understanding of both fundamental and practical aspects of the dynamical system behaviour in many different fields of physics (e. g. [1]). In particular, the behaviour of simple non-dissipative mechanical systems with few degrees of freedom presents an elementary case well suited for the study of the fundamental origins of dynamical randomness, with possible further extension to more complex situations. The description of chaos in such elementary dynamical systems within the formalism of classical mechanics has seemed to be rather successful and self-consistent [1-4]. At the same time its proposed quantum-mechanical versions, despite a large amount of the efforts made, have failed in creating a similar prosperous situation, even though a number of important particular results has been obtained [4-7]. The problem of the very existence of the truly unpredictable behaviour of deterministic quantum systems remains unsettled [8]. The fundamental difficulty stems from the unavoidable wave involvement in quantum postulates: waves do not easily show global instabilities necessary for the development of chaotic regimes or, in other terms, waves lead to discreteness (one may physically realise minimum a half-wave and not, say, one fifth of it), and the discreteness is incompatible with the existing notion of instability appealing to infinitesimal values and apparently indispensable for the known definitions of chaos. Solutions to this basic problem, proposed or implied, vary over a still wide but already shrinking range.

The most popular point of view involves the reduction of quantum chaos to some kind of very intricate but basically regular behaviour related eventually to the peculiarities of involved mathematical objects like zeta-functions [4,9] or random matrices [10]. Whatever the variations of such approach, it suffers from the evident conflict with the rather attractive idea, strongly supported by the existence of different types of classical chaos, that the world's complexity should be greater than zero: the latter would be impossible without the irreducible 'true' unpredictability and randomness in quantum mechanics as the most general description of the world. If it is not the case, i. e. if, for example, algorithmic complexity of quantum dynamics turns invariably to zero, then one possibility is that the existing quantum mechanics is no more general enough to provide a non-contradictory description for the world of chance [11].

The practical absence, however, of such properly chaotic scheme of quantum mechanics leaves the way open for another possibility: the conventional quantum mechanics itself is valid, for classically chaotic quantum systems in the same sense that for the classically regular ones, while the chaoticity of its classical limit 'appears' due to the particularities of the semiclassical transition. The emerging "quantum chaology" [12] finally also proposes rather a possibility than a solution: it is *suggested* why quantum and classical descriptions *need not* provide similar concepts of chaos, but it remains unclear why and *how exactly* the quantum-mechanical reality with complexity zero (absolute dynamical predictability and reversibility) passes to its truly random limiting case with non-zero complexity at the other end of the semiclassical transition. The latter should therefore 'produce' complexity 'from nothing' by a still unknown mechanism.

This gives us also a hint to the third and logically the last remaining possibility: there is no problem neither with the semiclassical transition, nor with the conventional quantum mechanics, but rather with our understanding of chaos and complexity in general, even in classical mechanics. It implies that in reality the 'pure' randomness and the corresponding complexity may not exist at all, and the apparent world of chance is just an illusory product of the ever wavering spirit... This choice, rather exotic as it may seem, can be provided, nonetheless, with a precise rational basis: it is stated simply that as the quantum-mechanical "diffusion localisation time" [13], after which a system starts deviating from apparently chaotic behaviour, becomes practically infinitely large for classical objects, one can never have a chance to achieve this regime of chaos suppression in the classical world, contrary to small quantum objects (this opinion was communicated to the author by Dr. S. Weigert from Basel University). The world is thus not complex, within this possibility, neither quantum-mechanically, nor classically, even though it is permitted to be very intricate. It represents, in fact, a huge generator of random numbers, basically deterministic but very perfect in its simulation of randomness (in other words, the world's dynamics is a periodic one, but the period is extremely large). In this way the problem of the origin of quantum chaos is effectively reduced to the alternative between the true complexity and a simple intricacy of being.

Apart from this logical choice among the three basic possibilities, one may mention the hypotheses about the quantum measurement involvement in chaos (see e. g. the special NATO Workshop Proceedings [14]) or the "fundamentally irreducible" representation by the density matrix [15]. It is not difficult to see that these approaches can also be eventually reduced to one of the above general possibilities touching the roots of the quantum paradigm. At their present state, however, they seem to be either too abstract, or too special to form a universal basis for the first-principle understanding of quantum chaos in real physical systems.[*)]

In the absence of clearness in the fundamental aspects, there is the general tendency to study particular features, or "signatures", of chaos in the *regular* quantum dynamics of the classically chaotic systems [6,7] leaving aside the related logical puzzles. However, it becomes more and more evident that one cannot hope to be really successful even in such particular research without the

---

[*)] We do not discuss here the recently appearing multiple attempts to relate deterministic randomness in quantum systems to a specific behaviour of the *dissipative* quantum systems inevitably involving an external source of irreversibility, whether it is stochasticity of the environment or the equivalent "coarse-graining" of a problem. We are looking for an intrinsic causal source of dynamic unpredictability which would remain even in the zero-noise limit, even though the addition of noise may play an important practical role in complex system behaviour (sections 2.3 and 5 of the full paper). This approach involves eventually considerable extension of the existing interpretation of *any* type of chaotic behaviour, including the "divergent-trajectories" paradigm of classical deterministic chaos (sections 6 and 10 of the full work) that suffers, in fact, from the same incompleteness as the "dissipative chaos" concept.

general self-consistent picture including the basic issues (not to mention the temptations of pure curiosity). This situation predetermines the importance of the search for a general self-consistent quantum (and eventually any other) chaos description providing the fundamental origin of randomness in deterministic systems. In this essay we present such an approach starting from the application of the unreduced version of the well-known optical potential method (see e. g. [16]) to the analysis of quantum chaos in Hamiltonian systems with periodic perturbation. This approach appeared originally as a part of quantum-mechanical description of charged particle scattering in crystals [17] revealing its chaotic behaviour (see, in particular, section 2.5 of the cited article). The results obtained are considerably developed and generalised in the present paper.

Using a generic example of arbitrary periodically perturbed Hamiltonian system, we show that our method naturally leads to the new concept of the fundamental origin of chaos in dynamical systems. In particular, it permits one to overcome the 'pathological regularity' of quantum mechanics and to perform the ordinary semiclassical transition also for chaotic systems, in agreement with the correspondence principle. It is worthwhile to note that our approach is presented in the form ready for its application to practical study of Hamiltonian quantum chaos in various real physical systems including both basic aspects and the particular analysis of the measured quantities.

A separate important topic concerning fractals involvement with chaos is considered in section 4 (of the full paper), where we show, within the same approach, that fractals naturally appear as solutions of the modified Schrödinger equation for a chaotic system that can be obtained analytically and not only by computer simulations (the result refers, in fact, to any dynamic equation presented in the proper form). In fact, we demonstrate that any solution describing a system in a chaotic regime (and this is the absolute majority of all situations) has a fractal character, and the properties of this fractal can be determined by the outlined procedure.

It is important that in order to obtain quantum dynamics with non-zero complexity and the conventional semiclassical transition, one does not need to reconsider the foundations of quantum mechanics as such but rather to use another, more general, form of the same formalism. This can help to moderate the painful choice described above (we discuss these issues in more detail in section 5 of the full paper). Moreover, in Part II of the full work (sections 8-10) we show that it is the main unsolved problems of the foundations of quantum mechanics, known as quantum indeterminacy and wave reduction, that can be given transparent causal solutions by application of the same method to the process of quantum measurement.

Finally, the purpose of this work is to demonstrate the universal character of the results obtained permitting one to extend the same concept of complex behaviour to other Hamiltonian and non-Hamiltonian, classical, and eventually distributed nonlinear systems. The issuing universal notions of the fundamental dynamic uncertainty, randomness, probability, complexity, (non)integrability, and general solution are introduced in section 6 (full version).

# 2. Quantum chaos as a manifestation of the fundamental dynamic uncertainty (≈ section 7 of the full paper)

A new method for the dynamical chaos analysis is presented by its detailed application to the description of quantum chaos in periodically perturbed Hamiltonian dynamical systems. We consider in parallel two versions of such a system: the time-independent conservative system consisting of a regular part and a periodic perturbation, and the time-dependent case with a regular part disturbed by a time-periodic potential. The results for the two versions are generally quite similar with the distinctions specified where necessary.

The key result of the method consists in the discovered additional plurality of the solutions of a dynamical equation (here the Schrödinger equation) presented in certain modified form, where the ordinary main dynamical function (here the potential) is replaced by the corresponding effective dynamical function (see, in particular, section 5 of the full paper). This modified form is obtained from the ordinary form by simple algebraic transformations, familiar from the optical potential formalism, which should give, a priori, the equivalent equation. However, the effective potential is found to be a multivalued function and the excessive number of solutions cannot be reduced to any known, or spurious, effect; it forms a set of 'realisations' depending on the type of a system and its parameters. Each realisation corresponds to the 'normal' complete set of solutions obtained for the corresponding 'branch' of the effective dynamical function and providing a possible value of each observable quantity. It is shown that if one accepts this modified form of the Schrödinger equation as the basic one, then the discovered existence of more than one realisation for a chaotic system can be interpreted as 'true' quantum chaos.

Indeed, quantum chaos described in this way involves the unreduced dynamical unpredictability as well as its transformation, in the semiclassical limit, to the corresponding results for the same system obtained within classical mechanics. This agreement with classical results includes a number of qualitative features of chaos (the existence of the regimes of global chaos and global regularity and their general character, like the asymptotically vanishing remnants of chaos in the domain of global regularity) and the quantitative expression for the point of transition chaos-regularity (called here the 'classical border of chaos', for the general quantum case).

At the same time a number of purely quantum features is also obtained within the proposed description. One of them is the "quantum suppression of chaos" which, contrary to the majority of the existing conceptions of quantum chaos, is generally only partial in our description, allowing of existence of the developed quantum chaos even far from the semiclassical limit. There are at least two common types of this quantum suppression of chaos.

The first one has a universal character and is due simply to the finite energy-level, and therefore realisation, separation in quantum case, which leads only to a partial disappearance of the manifestations of chaos.

The second one is more specific: it is observed in the case of Hamiltonian systems with time-independent periodic perturbation and is absent in the time-dependent case. It is shown that this kind of quantum suppression of chaos can be either partial, or complete depending on parameters. It becomes complete, i. e. a generally chaotic system transforms into a strictly regular one, when the full energy of the system is less than certain characteristic quantity called the 'quantum border of chaos' which is close to the lowest energy level.

This conclusion has the important physical consequences because it explains, in a remarkably simple way, the absolute stability of the elementary constituents of matter like atoms and nuclei, in their ground states, despite the existence of the developed quantum chaos regimes predicted above. However, the latter can also play an important role in the behaviour of the elementary quantum bound systems. It is clear, for example, that any excited state for such system is generically chaotic. Moreover, the existence of the specific chaotic ground states can also be expected.

It is shown thus that the quantum and classical borders of chaos and the dynamical regimes that they determine represent certain rather universal types of the chaotic system behaviour. It is important to emphasize, however, that according to the performed analysis, quantum chaos is compatible with the existence of many more general and particular variants of chaotic behaviour; they are determined by the set of the possible realisations for a chaotic system and their parameter dependence, which is illustrated by a number of examples. The proposed formalism permits one to reveal and study in detail these dynamic regimes for any particular quantum system.

One of the quantum chaotic regimes is especially interesting: it represents a kind of quantum chaos without any classical analogue. It appears in the form of large irregular 'quantum jumps' of the effective potential characteristics and, correspondingly, of the measured quantities. In particular, the height of a potential barrier can suddenly and considerably diminish; a barrier may effectively disappear at all, or it can even be transformed into a well, and vice versa. In the suitable experimental conditions this phenomenon may manifest itself as a version of the specific 'chaotic quantum tunnelling'.

The same method leads to the prediction of chaotic behaviour also for the free-motion states of quantum system possessing a transparent interpretation and a classical analogue.

Another feature of the chaotic quantum dynamics revealed within the formalism of the fundamental multivaluedness is its fractal character. The main dynamical function, the effective potential, is composed of many fractal branches, each of them corresponding to a realisation of a problem. The fractal smearing is superimposed on the fundamental multivaluedness and nonlocality of the effective dynamical function that is also referred to as the fundamental dynamical fractal of a problem. It is shown how the details of the chaotic dynamics can be translated in terms of the fractal geometry of this peculiar object, which outlines the way for the complete geometrisation of the complex (quantum) dynamics.

Generalising all the results above, we may say that they form a starting point of what could be called the 'quantum mechanics of chaotic systems', which is fundamentally different from the approach of the canonical "quantum chaos" whatever is the exact meaning of the latter. Indeed, we have seen that one does need to use another, extended, form of quantum formalism for a non-contradictory description of complex systems, and therefore chaos in quantum world is something more profound than another specific physical phenomenon; in fact, it cannot be separated from the fundamentals of quantum dynamics, but at the same time there is no any direct involvement with the primary quantum postulates. In other words, the revealed fundamental dynamic uncertainty is at least as 'big' as the quantum-mechanical indeterminacy (in Part II we show that the latter can, in fact, be reduced to the former). The logically natural construction of the modified quantum formalism of fundamental multivaluedness provides just a proper combination of the two uncertainties leading, as we have seen, to agreement with classical mechanics and to a variety of complex (in a 'good' sense! — see section 6 of the full paper) patterns of quantum behaviour. Further understanding and detailed investigation of this complex quantum world may constitute a subject of the new 'chaotic' quantum mechanics.

It is also demonstrated that the proposed method and the concept of the fundamental multivaluedness of dynamical functions can be, most probably, extended to other types of chaotic systems providing thus a universal basis for the complete description of their dynamics, and the ultimate origin of randomness and complexity. In particular, an invariant definition of physical complexity is proposed (section 6 of the full paper). It is based on the set of realisations for a system, naturally satisfies the general requirements for such definition and can be directly applied to real dynamical systems providing e. g. the dependence of their complexity on parameters. This universal concept of dynamic complexity involves also the rigorous transparent definitions, and extensions to arbitrary systems, of such notions as (non)integrability, effective nonlinearity, general solution (complete system of solutions), and probability (randomness).

The fundamental alternative concerning the complexity of the world (see section 1 above) is solved thus in favour of complexity, within our approach. Nonetheless the other possibility, outlined in the Introduction, also remains formally non-contradictory, even though to our opinion a number of Gedanken experiments could provide serious doubts in the viability of the zero-complexity world. The results of Part I of this work are effectively reduced to the discovery of a plausible modification of quantum mechanics with non-zero complexity, which provides effective support for these doubts about the absolutely calculable universe. Apart from doubts and hopes, however, neither of the two possibilities can be conclusively accepted or rejected at the present moment. Their further comparison needs additional theoretical and experimental work.

# 3. Incompleteness of quantum mechanics and the involvement of chaos (≈ section 8 of the full paper)

The ordinary scheme of quantum mechanics cannot provide non-zero complexity for Hamiltonian dynamics [11]; for any reasonable definition of complexity, this is equivalent to absolute dynamic predictability, reversibility, and the absence of the true chaos in quantum world. The issuing conflict with the predominantly chaotic behaviour of the counterpart classical systems (see e. g. [1-3]) leaves us with only two possibilities: either we have just an illusion of randomness and complexity in the form of classical chaos, while the world is basically predictable, or the standard quantum mechanics is not complete, at least for chaotic systems (see also section 1 above). At present, the choice within this alternative can be made rather by general considerations, largely non-physical and hardly rigorous. If one tends to accept that the highly inhomogeneous world, filled with extremely sophisticated spatiotemporal mixture of regularity and hazard, is more consistent with non-zero complexity, then he enters automatically into the *first level* of chaos involvement with the foundations of quantum mechanics. This position is well illustrated by the questions put in the title of an article on algorithmic complexity of quantum dynamics of classically chaotic non-dissipative systems (first paper in ref. [11]):

$$\text{Does quantum mechanics obey the correspondence principle ?}$$
$$\text{Is it complete ?} \qquad (1)$$

The definite (and negative) answers implied inevitably invoke, however, the third, hidden, question: is it the *same* incompleteness as that known before, existing from the birth of quantum mechanics and discovered by the Founding Fathers themselves? The rather positive answer of the authors of the cited paper can be certainly justified by the apparent awkwardness of the situation, where the fundamental incompleteness exists in several different types. However, as is shown in Part I (see section 2 above), one can propose certain natural modification of quantum formalism (represented by the Schrödinger equation) such that the correspondence principle is re-established in its conventional form also for chaotic systems. As this modified Schrödinger equation is obtained from its ordinary form by simple algebraic transformations and is reduced to it for regular dynamics, it inherits, a priori, all the basic axioms of the ordinary quantum mechanics including their incompleteness. At the same time we have no more contradictions, within this modified version, concerning the correspondence principle. Therefore the 'chaos-induced' incompleteness can be removed, in a natural and self-consistent manner, providing in addition a variety of chaotic quantum dynamic regimes characterised by the well-defined physical complexity. In other words, the problem of quantum chaos can be resolved within the introduced concept of the fundamental dynamic uncertainty (or fundamental multivaluedness of dynamical functions) which *does not depend* on the basic quantum postulates with all their strong and weak sides. These results permit one to advance in the positive direction while estimating the consistency of quantum mechanics (1):

$$\begin{array}{c} \text{Quantum mechanics (in the modified form)} \\ \text{obeys the correspondence principle.} \\ \text{Still it is incomplete.} \end{array} \qquad (2)$$

From the other hand, it may seem that in this way the new 'chaotic' quantum mechanics delays possible solution for the old problem of incompleteness and even, in a sense, plays for the latter by permitting it to survive even the hard trial of quantum deterministic randomness. However, in Part II we show that, in fact, it is quantum-mechanical indeterminism, representing the most mysterious part of the fundamental quantum problems, that depends on, and can be effectively reduced to, the fundamental dynamical uncertainty revealed in Part I. In this way the *formal* incompleteness of quantum mechanics can be practically removed. This forms *the second level* of chaos involvement with the foundations of quantum mechanics.

We note at once that the general idea about the relation between the two types of uncertainty, the dynamic and the quantum ones, seems to be rather evident and has already been exploited, from both ends (see e. g. [14]). However, these investigations only emphasised the conclusion following already from the general state of things in quantum chaos: one certainly could not reduce the true quantum indeterminacy[*)] to a basically regular regime of (conventional) quantum chaos. Indeed, it is clear that the question itself about chaos involvement with the irreducible quantum indeterminacy can seriously be posed only when one has the *true* quantum chaos providing a fundamental dynamical source of randomness. This is exactly the case for the concept of the dynamic multivaluedness presented in Part I, and in Part II we propose the detailed realisation of the appearing hope to obtain a causal solution for the problem of quantum measurement. It is important that the price paid for this solution, the postulate of the fundamental dynamic uncertainty, seems to correspond well to the obtained results, in every sense: by its profound basic nature, universality, novelty, and practical efficiency (see section 2 above).

This point of contact between wave mechanics and dynamic complexity could be anticipated not only from the side of deterministic randomness tending to find its place in quantum world, but also starting from intrinsic tendencies in understanding of basically irreducible roots of quantum indeterminacy. Indeed,

---

[*)] Note the difference between the fundamental property of unpredictability of detailed parameters of individual particle manifestation within the Schrödinger wave, which is referred to as *indeterminacy* in this paper, and the *uncertainty* of the observed wave characteristics entering Heisenberg's relations. Even if the two notions are not independent, the former concerns the fundamentals of the wave-particle dualism as they appear in the measurement process, while the latter concentrates more on the wave nature of quantum objects. In any case, within this paper we deal almost entirely with indeterminacy, and we prefer using this term to avoid any confusion, while leaving the term uncertainty for other related, but clearly distinct, concepts (e. g. the fundamental dynamic uncertainty of *any* chaotic dynamics which is shown to be the *eventual* origin of quantum indeterminacy, cf. quant-ph/9902015).

after 70 years of quite intensive research on the problem, it persists in its paradoxical status of something that practically works perfectly well in great variety of situations, but nevertheless cannot be understood at the level of the most general logic forming the base of the scientific method itself. The tensions created are so large that an appreciable part of quite serious approaches tends to admit the fundamental fail of this method, either by deducing that we deal with a basic cognition barrier (see e. g. [35] for the concept of the "veiled reality", *réel voilé*), or by replacing the escaping physical solutions with philosophic maxima (this was characteristic for the Bohr's position: "the task of physics is not to find how the nature is, but rather what can we say about it", see [36], p. 8), or even by linguistic arrangements (see e. g. [37]).

In this situation a practical agreement, the famous *Copenhagen interpretation*, is generally accepted like a standard providing the optimal *compromise* between the known and the unknown that is involved with the as much famous *complementarity* principle of Bohr. To our opinion such understanding of the "standard" interpretation corresponds exactly to the role it really plays in quantum mechanics and permits one to avoid both typical extreme cases of confusion, where what is in fact just a reasonable (and temporary) compromise is misunderstood either as the ultimate fundamental truth, or as a gross blunder with tragic consequences. In general, after the long absence of a really self-consistent and fundamental enough solution for quantum mysteries, there is a tendency to treat this or that evidently partial solution rather as the final one that excludes further crucial progress.

It is not really surprising to find out that the existing exceptions from this rule lead to most creative approaches, though often underestimated and even disregarded, like that of de Broglie, even if neither of them has provided, up to now, the final answers. Indeed, it is well-known that a future physical concept takes its shape by properly formulated *questions* issuing from the basically relevant, though often intuitive, *physical* thinking, rather than by immediate formal answers. This attitude, advocated so vigorously by Louis de Broglie and starting always from the direct but subtle contact between Reason and Nature, has proven to be the only one capable of unravelling any great mystery. The solution to the problem of quantum indeterminism proposed in Part II and its implications are also inspired by this *synthetic* approach, and we consider it to be not out of place to clearly designate this way of thinking as a general but indispensable base for the detailed constructions. It is quite natural that, as we shall see, the particular results obtained here are profoundly consistent with the causal quantum mechanics of de Broglie, which leads finally to a tentative global scheme of the complete quantum theory, *quantum field mechanics* (section 10 of the full paper).

Independent of these our findings, it seems to be rather evident that the 70-year search for the complete quantum mechanics approaches now its turning point. One indication of it is felt from the character of internal development of the totality of numerous "interpretations" whatever their relative popularity and conceptual weight within the whole field. Well illustrated by the recently appeared and differently oriented books [35,36,38][*], the current situation shows pronounced signs of basic saturation of principal point discussion both within each approach, and between them. From the other hand, the emerging new ideas of the "physics of complexity" (see sections 1, 2) are sufficiently fundamental and revolutionary in order to ensure, after being presented in the proper form, a crucial positive advance of the most realistic causal interpretations. Before specifying the latter possibility, we briefly outline some relevant details of the problem.

As was noticed by many researchers, the fundamental problems of quantum mechanics, and eventually the proposed postulates, fall naturally apart into two related but still distinct, and even opposed, components. This basic duality can be seen from different points of view, and the earliest and most general formulation is reduced simply to the fundamental wave-particle dualism. Indeed, if we take experimentally evident wave nature of micro-objects as an axiom, then the equally evident localised particle manifestations of the same objects are not only difficult to explain, but even taken as axiom they contradict the first one. While the practical-purpose standard interpretation gets out of it by taking, in fact, this contradiction as the main axiom, a consistent solution to this basic problem is claimed to be found within the approach of D. Bohm proposed in 1952 and presented in the most complete form in [36]. In this interpretation the coexistence of waves and particles, both real, is postulated from the beginning (though their detailed physical origin is *not* specified from the first principles, contrary to the complete version of de Broglie approach); of course, the two entities are coupled, and the proper choice of coupling seems to reconcile wave and particle aspects. However, the God appears to be more subtle than this. As various experiments show, not only particle and wave coexist, but in each its manifestation the particle 'kills' its wave without leaving any trace of it; the latter seems to be instantaneously shrinking around the particle showing itself in interaction with external objects: this is the famous "wave reduction". And as if it is not enough to bury any good theory, it appears that particle commits this unfaithful act in a provocatively irregular fashion: one can only know its probability depending on place and time, but never exactly where and when. Both these features cannot be intrinsically incorporated in the existing scheme of Bohm's causal interpretation: like so many other approaches it is finally forced to accept indeterminacy without fundamental physical substantiation, but rather in the form of a usual semiempirical probability definition, and to regard the reduction as a plausible consequence of omnipresent environmental influences [36]. The real, and positive, role of Bohmian mechanics is that it serves as the optimal, though not complete, reconciliation of a *mathematically* consistent description with the *physically* consistent causal *double-solution (pilot-wave)* interpretation of de Broglie conceived and further developed during practically all the period since the appearance of quantum mechanics (see e. g. [41-43]). We have here generally the same type of relation between Bohm's theory and the original de Broglie approach as that between the standard Copenhagen interpretation and (complete) quantum mechanics in the whole, and it should be estimated with the same comprehension mentioned above (note that the so-called 'pilot-wave' interpretation, equivalent to the Bohmian mechanics, was proposed by de Broglie himself, 25 years before Bohm, as an explicitly simplified 'model' for the underlying complete version of the 'double solution', see e. g. e-print quant-ph/9902015).

---

[*] The excellent selections of other references on the subject can be found therein; recent review articles can be exemplified by refs. [39,40].

Thus the most mysterious among the observed quantum mechanical features are related to particle manifestations in physical interactions of its wave-pilot and comprise (wave) reduction and (particle emergence) uncertainty. These are the constituents of what is called quantum measurement, or rather the problem of measurement which resists to all attempts of consistent physical solution, as was explained above. Note that these difficulties do not directly touch the postulate about the wave implication and its dynamics as such. This leads us to another form of the above-mentioned duality of quantum problems appearing now as the couple 'wave dynamics - measurement process'. The first component in this couple does not generally invoke physical mysteries. In the usual scheme the wave dynamics is described by the Schrödinger equation (modified, according to the results of Part I, for chaotic systems). In the much more complete approach of de Broglie [41-43] it is the nonlinear wave dynamics which is supposed to give the "double solution", composed of the quasi-linear Schrödinger $\psi$-function and the localised highly nonlinear soliton-like[*] singularity ('particle'), these two parts forming a unique object. And although the precise mathematical description of this complex dynamics has never been found, these *technical* difficulties seem not to hide any *physical* puzzle. It is quite the contrary for the second dual component, the measurement process. The latter seems to be incomprehensible in terms of ordinary causal physics, and it is this contradiction which is at the origin of all the attempts to assume a 'particular quantum physics' and even a particular philosophy of quantum world. R. Penrose [38] makes a clear distinction between the paradoxes coming from quantum measurement, called **X**-mysteries, and unusual nonlocal (eventually, wave-related) quantum effects, **Z**-mysteries. In his remarkably precise general analysis he emphasises that the **Z**-mysteries can be understood, in principle, already within the existing physical and philosophical notions, though maybe not without considerable efforts, whereas the **X**-mysteries are basically incomprehensible within the current concepts, whatever the efforts, and need 'something else', some really new and fundamental notion-phenomenon-postulate(s), to be understood at the same level as other physical concepts.

---

[*] We use the terms 'soliton-like' (solution) and 'quasi-soliton' (or 'virtual soliton') for this peculiar part of the de Broglie double solution which was originally named "bunched wave" (*onde à bosse*), and later on referred to also as "singular wave" (*onde singulière*). We emphasize the difference of this object with respect to classical solitons known as exact solutions of certain nonlinear equations. In particular, quasi-soliton *must not* be an exact solution; it is rather an unstable (chaotic) solution of a nonlinear equation. We still use the word 'soliton' in its designation referring to the property to form a high abrupt hump, though unstable.

We propose such a solution for the problem of quantum measurement (section 9 of the full paper). We shall see that it corresponds to the natural physical type of approach described above, being expressed, at the same time, in an unambiguous mathematical form. It is inevitably based on a new fundamental concept, that of the fundamental dynamic uncertainty, introduced in Part I and extended here to the case of quantum measurement. This solution does not directly concern the first part of the dual quantum problem, the wave postulate (even though it basically permits of existence of a *material* wave), but it is especially consistent with the causal physical picture of the double solution of de Broglie and effectively provides the real hope to specify and extend its formulation (section 10 of the full paper). In other words, we do try to 'cross off' the **X**-mysteries from the list of quantum mysteries, as it has been anticipated [38], to be left only with tractable (and now *easier* tractable), though non-trivial, **Z**-mysteries of effectively nonlinear wave mechanics of the real material field(s).

# 4. Quantum mechanics with complexity and the advent of wholeness (≈ the end of section 10 of the full paper)

We conclude this report with the emphasis on the role of complexity, the latter being always understood in the same well-defined sense (section 6 of the full paper), in the proposed causal explanation for quantum indeterminacy and wave reduction. We argue that this implication of deterministic randomness in the resolution of the most puzzling quantum **X**-mysteries is inevitable for at least two complementary reasons. First, the resolution of such basic problems of the wave behaviour could only be possible at the expense of a new concept, not less fundamental than the wave postulate itself (see also [38]), and it is not easy to imagine another candidate for this role, apart from a universal postulate introducing dynamic complexity. Second, the dynamic chaos itself should certainly find its place in quantum mechanics describing the complex world, and this place can only be one of the basic ones. The fact that up to now the true chaos seemed to be incompatible with quantum mechanics just shows, as we have seen, that the standard quantum formalism is not adapted to the interpretation of complexity. It is, by the way, the case of any nonlocal approach including that of Liouville equation in classical mechanics. We have demonstrated, however, that already simple algebraic transformations of the ordinary formalism lead to the appearance of the fundamental dynamic uncertainty that can manifest itself as the 'ordinary' quantum chaos or as the fundamental quantum indeterminacy. From the other hand, the same quantum indeterminacy should become evident, and intuitively more transparent, in the anticipated local formulation of quantum field mechanics, but this demands the construction of the local nonlinear formalism, in continuation of de Broglie ideas. Now, however, the latter task seems to be much more feasible: in Parts I and II (of the full work) we reveal different versions of a universal mechanism showing how the *effective nonlinearity*, being a synonym of complexity and really existing in the physical world, can be put into a natural explicit form just by the properly presented formalism, without any artificial constructions. Once appeared after the long and vain search for it, the nonlinearity of wave mechanics will certainly give us a variety of the known, anticipated, and now inconceivable possibilities.

    It would not be out of place to recall that the founders of the Copenhagen interpretation, led by Niels Bohr, had seen the final victory of their approach in the definite exclusion from quantum physics of the ordinary 'macroscopic' intuition, based on everyday experience and especially on its 'fuzzy' human logic. Now, seventy years after they have won, it is precisely this type of logic that reappears as the non-contradictory causal scheme of wave mechanics within the described synthesis between the defeated 'fuzzy' approach of de Broglie and the universal concept of dynamical complexity. We have seen that almost humanly intricate, unpredictable and multiform, behaviour of the effectively nonlinear material wave becomes quite natural, and even inevitable, provided a 'gentle' (and mathematically correct) modification of the basic formalism is accepted in

exchange for the complexity of the world. This 'humanization' of quantum mechanics has been anticipated as one of the irreducible constituents of the beginning, and unavoidable, fundamental return of wholeness into the entire system of knowledge.

The involvement of complexity at the very basis of quantum mechanics is profoundly related also to the universal hierarchical structure of the World. Indeed, the most fundamental level of description of the complex world should certainly contain dynamic complexity in explicit form. This demand is now satisfied for quantum mechanics within the formalism of dynamical multivaluedness. From the other hand, one should be able to obtain complexity at any higher level of description, e. g. in classical mechanics, in distributed system behaviour, etc., without leaving that level. It is extremely important, that it is the *same* mechanism of the fundamental dynamic uncertainty that provides complexity (chaos) at each level, though with some particular details specific for that level (section 6 of Part I). In particular, and this is symptomatic, the formalism applied to reveal the manifestation of dynamic uncertainty in the measurement process, can be used with only minor changes for the description of complex behaviour in dynamic systems from a very large class (it is sufficient to consider the measured object and the instrument as abstract interacting physical systems). It means that we have the *double correspondence* between different levels of complexity, within our approach: the *direct* one, where the results at a higher level can be deduced from the more fundamental description (e. g. the results for classical chaotic systems can be obtained within the purely quantum-mechanical consideration, see sections 2.3, 3 of Part I); and the *conceptual* correspondence, where the complexity at a higher level can be obtained without any reference to the underlying fundamental level, but within the same concept and method that the complexity at that lower level. This 'vertical' double correspondence, accompanied with the 'horizontal' global correspondence principle (section 10 of the full paper), provides another evidence in favour of a self-consistent holistic picture of the Complex World outlined throughout the present work (see also physics/9806002).

We can give finally a well-substantiated positive answer to the basic questions (1) considerably extending our preliminary answer (2) and outlining a feasible issue towards the physically and formally complete quantum field mechanics:

$$\begin{array}{r}\text{Quantum mechanics (in the modified form) obeys}\\ \text{the (global) correspondence principle.}\\ \text{It is formally complete, but physically incomplete.} \quad (3)\\ \text{It can be extended by the addition of the local nonlinear theory.}\end{array}$$

# References

[1]  E. Ott, *Chaos in dynamical systems* (Cambridge Univ. Press, Cambridge, 1993).
[2]  A.J. Lichtenbegr and M.A. Lieberman, *Regular and Stochastic Motion* (Springer-Verlag, New-York, 1983).
[3]  G.M. Zaslavsky, *Chaos in dynamical systems* (Harwood Academic Publishers, London, 1985).
[4]  M.C. Gutzwiller, *Chaos in Classical and Quantum Mechanics* (Springer-Verlag, New York, 1990).
[5]  B. Eckhardt, Phys. Rep. **163**, 205 (1988).
[6]  *Chaos and Quantum Physics*, edited by M.J. Giannoni, A. Voros, and J. Zinn-Justin (North-Holland, Amsterdam, 1991).
[7]  F. Haake, *Quantum signatures of chaos* (Springer, Berlin, 1991).
[8]  *Quantum and Chaos: How Incompatible ?* Proceedings of the 5th Yukawa International Seminar (Kyoto, August 1993), Ed. K. Ikeda, Progress Theor. Phys. Suppl. No. 116 (1994); W.-M. Zhang and D. H. Feng, Phys. Rep. **252**, 1 (1995).
[9]  E.B. Bogomolny, Nonlinearity **5**, 805 (1992).
[10] O. Bohigas, in *Chaos and Quantum Physics* (ref. [6]); O. Bohigas, S. Tomsovic, and D. Ullmo, Phys. Rep. **223** (2), 45 (1993).
[11] J. Ford and G. Mantica, Am. J. Phys. **60**, 1086 (1992); see also J. Ford, G. Mantica, and G. H. Ristow, Physica D **50**, 493 (1991); J. Ford and M. Ilg, Phys. Rev. A **45**, 6165 (1992); J. Ford, in *Directions in Chaos*, edited by Hao Bai-lin (World Scientific, Singapore, 1987), Vol. 1, p. 1; J. Ford, ibid., p. 129.
[12] M.V. Berry, in *Chaos and Quantum Physics* (ref. [6]).
[13] B.V. Chirikov, in *Chaos and Quantum Physics* (ref. [6]).
[14] *Quantum Chaos - Quantum Measurement* (Kluwer, Dortrecht, 1992).
[15] I. Prigogine, Physics Reports **219**, 93 (1992).
[16] P.H. Dederichs, Solid State Phys. **27**, 136 (1972).
[17] A.P. Kirilyuk, Nuclear Instrum. Meth. **B69**, 200 (1992).
[18] A.I. Akhiezer, V.I. Truten', and N.F. Shul'ga, Phys. Rep. **203**, 289 (1991).
[19] G. Casati, I. Guarneri, and D. Shepelyansky, Physica A **163**, 205 (1990).
[20] B.V. Chirikov, Phys. Rep. **52**, 263 (1979); see also ref. [13].
[21] G. Hose and H.S. Taylor, J. Chem. Phys. **76**, 5356 (1982).
[22] G. Hose, H.S. Taylor, and A. Tip, J. Phys. A **17**, 1203 (1984).
[23] S. Adachi, M. Toda, and K. Ikeda, Phys. Rev. Lett. **61**, 655, 659 (1988).
[24] T. Dittrich and R. Graham, Europhys. Lett. **7**, 287 (1988).
[25] G.M. Zaslavsky, R.Z. Sagdeev, D.A. Usikov and A.A. Chernikov, *Weak chaos and quasi-regular patterns* (Cambridge Univ. Press, Cambridge, 1991).
[26] N.W. Ashcroft and N.D. Mermin, *Solid State Physics* (Holt, Rinehart and Winston, New York, 1976).
[27] S. Tomsovic and D. Ullmo, Phys. Rev. E **50**, 145 (1994).
[28] H.-O. Peintgen, H. Jürgens, and D. Saupe, *Chaos and Fractals . New Frontiers of Science* (Springer-Verlag, New-York, 1992).

[29] S.R. Jain, Phys. Rev. Lett. **70**, 3553 (1993); I. Guarneri and G. Mantica, Phys. Rev. Lett. **73**, 3379 (1994).
[30] J. Feder, *Fractals* (Plenum Press, New York, 1988).
[31] G. Jona-Lasinio, C. Presilla and F. Capasso, Phys. Rev. Lett. **68**, 2269 (1992).
[32] G. D'Alessandro, A. Politi, Phys. Rev. Lett. **64**, 1609 (1990).
[33] A. Crisanti, M. Falcioni, G. Mantica, and A. Vulpiani, Phys. Rev. Lett. **50**, 1959 (1994).
[34] I. Prigogine, *From Being to Becoming* (Freeman, San Francisco,1980); Can. J. Phys. **68**, 670 (1990); G. Nicolis and I. Prigogine, *Exploring Complexity* (Freeman, San Francisco, 1989).
H. Haken, *Advanced Synergetics* (Springer, Berlin, 1983).
[35] B. d'Espagnat, *Le réel voilé* (Fayard, Paris, 1994).
[36] P.R. Holland, *The Quantum Theory of Motion* (Cambridge University Press, Cambrige, 1995).
[37] H. Wimmel, Il Nuovo Cimento **109 B**, 1065 (1994).
[38] R. Penrose, *Shadows of the mind* (Oxford University Press, New York,1994).
[39] M. Combourieu and H. Rauch, Found. Phys. **22**, 1403 (1992).
[40] C. Dewdney, G. Horton, M.M. Lam, Z. Malik, and M. Schmidt, Found. Phys. **22**, 1217 (1992).
[41] L. de Broglie, *Une Tentative d'Interprétation Causale et Non-Linéaire de la Mécanique Ondulatoire* (Gauthier-Villars, Paris, 1956).
[42] L. de Broglie, *La réinterprétation de la mécanique ondulatoire* (Gauthier-Villars, Paris, 1968); *Recherches d'un Demi-Siècle* (Albin Michel, Paris, 1976).
[43] *Louis de Broglie. Un Itinéraire Scientifique.* Ed. Georges Lochak (La Découverte, Paris, 1987).
[44] L.D. Landau and E.M. Lifshitz, *Quantum Mechanics* (Nauka, Moscow, 1989).
[45] L. de Broglie, *La théorie de la mesure en mécanique ondulatoire (interprétation usuelle et interprétation causale)* (Gauthier-Villars, Paris, 1968).
[46] S. Weigert, Phys. Rev. **A43**, 6597 (1991).
[47] R. Omnès, Annals of Physics **201**, 354 (1990).
[48] F. Fer, *Guidage des particules, ondes singulières*, in: *Louis de Broglie. Sa conception du monde physique* (Gauthier-Villars, Paris, 1973), p. 279.
[49] J.-P. Vigier, Found. Phys. **21**, 125 (1991).
[50] T. Waite, Ann. Fond. Louis de Broglie **20**, 427 (1996).
[51] L. de Broglie, *La thermodynamique de la particule isolée (thermodynamique cachée des particules)* (Gauthier-Villars, Paris, 1964).
[52] G. Lochak, *La thermodynamique cachée des particules*, in: *Louis de Broglie. Sa conception du monde physique* (Gauthier-Villars, Paris, 1973), p. 325.
[53] H. Everett, Rev. Mod. Phys. **29**, 454 (1957).
[54] D. Home and M.A.B. Whitaker, Phys. Reports **210**, 223 (1992).
[55] R.B. Griffiths, Phys. Rev. Lett. **70**, 2201 (1993).
[56] E.P. Wigner, in *The scientist speculates*, ed. by I.J. Good (Heinemann, London, 1961).